\documentclass[journal]{IEEEtran}

\AtBeginDocument{%
  }

\usepackage{multirow}
\usepackage{colortbl}
\usepackage{framed}
\usepackage{mdframed}
\usepackage{xcolor}
\usepackage{float}
\usepackage{bbding}
\usepackage{subcaption}
\usepackage{xspace}
\usepackage{algorithm}
\usepackage{algorithmic}
\usepackage{hyperref}
\usepackage{tabularx}
\usepackage{amssymb}
\usepackage{amsmath}
\usepackage{booktabs}
\usepackage[most]{tcolorbox}

\definecolor{findingBG}{gray}{0.95}

\newtcolorbox{findingbox}{
  enhanced,
  breakable,
  width=\linewidth,
  colback=findingBG,
  colframe=findingBG,
  boxrule=0pt,
  arc=0pt,
  outer arc=0pt,
  boxsep=0pt,
  left=5pt,
  right=5pt,
  top=4pt,
  bottom=4pt,
  before skip=0.6\baselineskip,
  after skip=0.6\baselineskip,
  enlarge left by=0pt,
  enlarge right by=0pt
}

\hypersetup{
    colorlinks=true,
    linkcolor=black,
    filecolor=black,      
    urlcolor=black,
    citecolor=black,
}

\newcommand{\tool}{\textsc{Axiom}\xspace}
\newcommand{\bench}{\textsc{AxiomBench}\xspace}

\newcommand{\stagea}{quality-constrained semantic perturbation\xspace}
\newcommand{\Stagea}{Quality-constrained semantic perturbation\xspace}
\newcommand{\STAGEa}{Quality-Constrained Semantic Perturbation\xspace}

\newcommand{\Stageaa}{Exemplar augmentation\xspace}
\newcommand{\STAGEaa}{Exemplar Augmentation\xspace}

\newcommand{\Stageab}{Deteriorative perturbation\xspace}
\newcommand{\STAGEab}{Deteriorative Perturbation\xspace}

\newcommand{\Stageac}{Context disruption\xspace}
\newcommand{\STAGEac}{Context Disruption\xspace}

\newcommand{\stageb}{hybrid quality calibration\xspace}
\newcommand{\Stageb}{Hybrid quality calibration\xspace}
\newcommand{\STAGEb}{Hybrid Quality Calibration\xspace}

\newcommand{\STAGEba}{Representative Candidate Selection\xspace}

\newcommand{\stagebb}{hybrid score calibration\xspace}

\newcommand{\STAGEbb}{Hybrid Score Calibration\xspace}

\definecolor{findingBG}{gray}{0.95}
\colorlet{shadecolor}{findingBG}

\begin{document}

\title{Benchmarking Code LLM-as-a-Judge via Constrained Perturbation and Calibration}
% \title{\tool: Benchmarking LLM-as-a-Judge for Code via Rule-Based Perturbation and Multisource Quality Calibration}

\author{Ruiqi Wang, Xinchen Wang, Cuiyun Gao, Chun Yong Chong, Xin Xia, and Qing Liao%
\thanks{Ruiqi Wang, Xinchen Wang, Cuiyun Gao, and Qing Liao are with Harbin Institute of Technology, Shenzhen, China (e-mail: 24s151158@stu.hit.edu.cn; 200111115@stu.hit.edu.cn; gaocuiyun@hit.edu.cn; liaoqing@hit.edu.cn). Cuiyun Gao is the corresponding author.}%
\thanks{Chun Yong Chong is with Monash University Malaysia, Bandar Sunway, Malaysia (e-mail: chunyong@ieee.org).}%
\thanks{Xin Xia is with Zhejiang University, Hangzhou, China (e-mail: xin.xia@acm.org).}%
}

% \renewcommand{\shortauthors}{Wang et al.}

% <concept_id>10011007</concept_id>
% <concept_desc>Software and its engineering</concept_desc>
% <concept_id>10010147.10010178</concept_id>
% <concept_desc>Computing methodologies~Artificial intelligence</concept_desc>

% \ccsdesc[500]{Software and its engineering}
% \ccsdesc[500]{Computing methodologies~Artificial intelligence}

% \keywords{code evaluation, large language models, data synthesis}


\maketitle
\begin{abstract}
Code from LLM-assisted development may contain defects like API hallucinations, motivating the development of code evaluation metrics to assess such code before deployment. As conventional rule-based metrics merely score code based on  similarity to reference programs, and execution-based metrics require expensive manual preparation for tests, researchers have developed LLM-as-a-judge metrics for both behavioral and non-behavioral code evaluation, and curated multiple benchmarks to validate their effectiveness. However, such benchmarks suffer from critical limitations: Some feature coarse-grained binary labels of passing/failing unit tests, unable to distinguish between fundamental and minor flaws; others utilize fine-grained criteria with manually annotated scores as ground-truth, but such criteria are often vaguely defined, introducing subjectivity and unreliability in human scores. Furthermore, they often directly prompt LLMs to generate code for evaluation, which may lead to unbalanced code quality distributions, failing to test the metrics' discriminative capability across varying quality levels.

To curate a diverse benchmark with a balanced program quality distribution and streamline manual annotation, we propose \textbf{\tool}, a novel perturbation-based framework. It reframes program scores as the refinement effort needed for deployment, with two stages: (1) {\bf\Stagea}, prompting LLMs to apply sequences of predefined perturbation rules to existing programs, precisely controlling each program's target quality to achieve a balanced distribution. (2) {\bf\Stageb}, selecting semantically distinct programs for benchmark diversity, and performing streamlined manual annotation with multisource context. With \tool, we curate \textbf{\bench}, a large-scale multilingual benchmark with 1,962 code evaluation problems in a more balanced score distribution and more reliable annotated scores than previous benchmarks. We further conduct an empirical study, revealing that while LLMs offer competitive performance on ranking programs and verifiable evaluation subtasks, they suffer from systematic pessimism bias and sample-level agentic instability, and remain unreliable for subjective, non-behavioral code evaluation.
% current LLM-as-a-judge metrics, while powerful at relative ranking, suffer from issues such as hallucinating flaws in functionality and code quality, and high score variance with complex agentic setups.

\end{abstract}
\section{Introduction}

Recently, large language models (LLMs) have been widely deployed in software engineering. With comprehensive pre- and post-training, these colossal models possess strong reasoning \cite{DBLP:journals/corr/abs-2412-16720,DBLP:journals/corr/abs-2501-12948} and tool use \cite{DBLP:journals/csur/QinHLCDCZZHXHFSWQTZLSXZ25} capabilities, enabling them to assist in maintaining complex repositories. Besides powerful LLMs like Google's Gemini 2.5 \cite{DBLP:journals/corr/abs-2507-06261} and OpenAI's GPT-4o \cite{DBLP:journals/corr/abs-2410-21276}, the industry has developed agentic applications like Cursor and Claude Code. However, code from AI-assisted development may be defective, with issues like API hallucinations \cite{DBLP:conf/sigsoft/ChenCGJLM25} and social bias \cite{DBLP:conf/aaai/LingRW025}, potentially leading to functional incorrectness or suboptimal quality. Therefore, evaluating such code remains crucial before deployment, and researchers have developed various metrics to quantify the code's overall quality.

While execution accuracy % (Pass@\(k\) in code generation benchmarks) \cite{DBLP:journals/corr/abs-2107-03374} 
is the standard metric for validating program behavior, it requires manually constructing environments and test cases, which is labor-intensive. Rule-based metrics like CodeBLEU \cite{DBLP:journals/corr/abs-2009-10297} and RUBY \cite{DBLP:conf/iwpc/TranTNNN19} compare programs against reference programs. While efficient to compute, they merely measure superficial lexical similarity. Manual approaches require human experts to thoroughly review the code, which is expensive and demands high levels of domain understanding. To reduce human reliance, researchers propose LLM-as-a-judge metrics for code evaluation, leveraging LLMs' rich domain knowledge and instruction-following ability, bypassing the need for reference programs and test cases.

Naturally, researchers curate code evaluation benchmarks \cite{DBLP:journals/pacmse/WangGGFCX25,DBLP:journals/corr/abs-2507-10535,DBLP:conf/coling/ZhaoLT0YL025} to verify whether LLM-as-a-judge metrics truly offer human-level code evaluation. These benchmarks typically provide requirements and programs for the metrics, and compare metric scores against human or test scores. However, they suffer from major limitations, hindering comprehensive assessments: 
(1) \textbf{Unbalanced quality distribution.}
% and inefficient curation}. 
While open-source programs are often used as ground-truths in code generation benchmarks \cite{DBLP:conf/kbse/FengLGCWG024}, they are unsuitable for code evaluation benchmarks, being highly skewed towards higher quality. Therefore, existing benchmarks \cite{DBLP:journals/corr/abs-2504-13472,DBLP:journals/jss/EvtikhievBSB23,DBLP:journals/corr/abs-2507-10535} use LLM-generated programs and label their scores with humans or tests. However, the generation phase often prompts LLMs without constraints \cite{DBLP:journals/pacmse/WangGGFCX25,DBLP:journals/corr/abs-2504-13472}, suffering from the same higher-quality skew and failing to adequately represent lower quality levels.
(2) \textbf{Coarse-grained or subjective evaluation criteria.} Some benchmarks use binary labels (pass/fail) to indicate functional correctness \cite{DBLP:journals/corr/abs-2502-12468,DBLP:conf/emnlp/TongZ24}, disregarding non-behavioral code quality and partial correctness; others apply fine-grained criteria \cite{DBLP:conf/eacl/Zhuo24,DBLP:journals/corr/abs-2504-13472} like a 5-point Likert scale, but with vague descriptors like ``slightly helpful'', instead of detailed and objective definitions. This vagueness introduces high subjectivity and inconsistency in human annotation, making the supposed ``ground-truth'' unreliable. 
(3) \textbf{Costly manual workload.} Fine-grained evaluation requires human-annotated scores, which is prohibitively expensive and prone to quality control issues. This stems from its immense cognitive load on annotators, requiring deep domain knowledge to meticulously reason about logic and dependency at scale. Thus, existing benchmarks are often small, unbalanced, and annotated with unreliable scores, highlighting the need for a novel method to curate a large, balanced, and reliably-grounded benchmark for meta-evaluating LLM-as-a-judge metrics.

% \footnote{AXIOM: {\bf A}ssessment via \textbf{X} (cross)-ver\textbf{I}fication and \textbf{O}racle-guided \textbf{M}utation. Although ``mutation'' is used here, we use ``perturbation'' for the remainder of this paper.}
Therefore, we propose {\bf \tool}, a novel framework to synthesize code evaluation benchmarks. \tool mitigates the shortcomings of existing scoring systems by reframing code evaluation into measuring the hypothetical effort needed to repair programs for deployment, resulting in fine-grained, interpretable scores where higher values signify more polished and deployable programs. By clearly defining each possible score based on the scope of required fixes (e.g., a localized tweak or a structural refactor), our evaluation criteria provide the necessary foundation for a robust two-stage methodology: (1) {\bf\Stagea}. Instead of prompting LLMs to generate code from scratch without constraints, we utilize a perturbation-based pipeline. % that requires LLMs to modify existing programs. 
This pipeline invokes LLMs to apply a sequence of predefined rules to alter a program's functionality or code quality, precisely generating programs of predetermined quality. The involvement of LLMs allows complex semantic changes to code, beyond simple, static tool-based changes like operator swapping. %, resulting in a large, diverse and balanced benchmark. 
Each program's functionality is checked via test cases. (2) {\bf\Stageb}. Due to the sheer number of programs generated, we select a subset of representatives by greedily choosing semantically distinct candidates to ensure diversity. We then perform a streamlined annotation: instead of functionality evaluation involving complex logic reasoning, the human annotator simply determines the code quality and scope of changes to label the final score. This greatly reduces manual workload, resulting in a large, balanced, verified benchmark even with limited human resources. 

We curate \bench with \tool, a code evaluation benchmark with 1,962 programs in a balanced quality distribution with reliable human-calibrated scores, and study existing LLM-as-a-judge metrics, making the following discoveries:

\begin{itemize}
    \item LLM-as-a-judge metrics are decent at relatively ranking programs based on quality, while agentic approaches further boost this capability for stronger LLMs. However, they suffer from systematic pessimism bias, rendering absolute scores unreliable without calibration.

    \item LLM-as-a-judge shows competence in verifiable tasks like functionality evaluation. Surprisingly, conventional rule-based metrics outperform LLMs in non-functional assessment. Agentic approaches only boost performance on verifiable tasks of moderate difficulty.

    \item Simpler metrics exhibit high sample-level consistency across repeated runs. In contrast, complex agentic workflows accumulate stochasticity across planning and tool-use steps, causing less stable scores for individual programs. However, their aggregate benchmark performance varies only marginally across full-dataset reruns.
    % \item Performance of LLM-as-a-judge metrics reveals a trade-off between LLM capability and metric complexity, and is constrained by systematic biases, preventing full human alignment.

    % \item LLM-as-a-judge metrics often hallucinate flaws in functionality and code quality, and cannot reliably estimate refinement effort by distinguishing tweaks from refactors.

    % \item Scoring consistency of LLM-as-a-judge metrics inversely relates to its procedural complexity, where complex agentic frameworks suffer more from notable scoring variance.
\end{itemize}

Our contributions can be summarized as follows:

\begin{itemize}
    \item We present \tool, a framework for generating code evaluation benchmarks at scale, to mitigate unreliability of previous data curation and annotation procedures.
    \item We design the \stagea stage, semantically perturbing code under strict score ceilings with LLMs to precisely control the expected quality. Predetermined quality scores are accurate, requiring substantially reduced manual workload for calibration.
    % We design a two-stage procedure for \tool. The first stage introduces advanced and varied semantic perturbations to programs, while precisely controlling the program quality distribution. The second stage proposes a streamlined manual workflow to calibrate the quality scores with multisource information, enabling efficient and accurate annotation.
    %, where the first guides LLMs to perform program perturbation to generate diverse data, whereas the second calibrates the scores using a multisource approach with reduced human effort.
    \item We release \bench and extensively study recent LLM-as-a-judge metrics, uncovering systematic biases and identifying key scenarios determining their reliability.
\end{itemize}

% The rest of this paper is organized as follows: Section \ref{sec:work} introduces work related to code generation, evaluation, and transformation. Section \ref{sec:method} describes the methodology of \tool. Section \ref{sec:setup} presents the statistics of \bench and the study setup. Section \ref{sec:res} analyzes the results with our benchmark. Section \ref{sec:dis} provides additional discussion. Section \ref{sec:con} concludes the paper.
\section{Related Work\label{sec:work}}
% \subsection{LLMs for Code Generation}
% Many LLMs are specifically trained for code-related tasks. %, sometimes referred to as large code models (LCMs). % CodeT5+ \cite{DBLP:conf/emnlp/WangLGB0H23} is an early attempt at instruction-tuning LCMs. % to comply with natural language instructions. 
% Code Llama \cite{DBLP:journals/corr/abs-2308-12950} is a family of 7-70B LLMs trained on 500B tokens, improving performance on benchmarks like HumanEval \cite{DBLP:journals/corr/abs-2107-03374}. Qwen3-Coder \cite{DBLP:journals/corr/abs-2505-09388} is a 480B Mixture-of-Expert (MoE) LLM with state-of-the-art performance on agentic coding and tool use. % on difficult benchmarks and supporting a context length of 1M.
% Meanwhile, many general-purpose LLMs also % are also developed to 
% deliver exceptional coding performance. % , especially with the internal reasoning capability, thanks to inference scaling \cite{DBLP:conf/iclr/Snell0XK25}. 
% DeepSeek-V3.1 \cite{DBLP:journals/corr/abs-2412-19437} is an open-source LLM supporting both non-reasoning and reasoning mode, with higher token efficiency % and agentic performance compared to 
% than its predecessor DeepSeek-R1 \cite{DBLP:journals/corr/abs-2501-12948}.
% Gemini 2.5 \cite{DBLP:journals/corr/abs-2507-06261} is a large multimodal model family from Google with state-of-the-art coding and reasoning capability. % in complex scenarios. %, equipped with multi-step acting and tool use capabilities.

\subsection{Code Evaluation Metrics}
% Execution accuracy \cite{DBLP:journals/corr/abs-2107-03374} is the standard metric to validate the functional correctness of the code. While tests can be executed automatically, constructing testing environments and curating unit tests for large numbers of coding problems demand tremendous manual effort.

Rule-based metrics measure lexical similarities between programs and reference programs with predefined rules. % While conventional NLP metrics like BLEU \cite{DBLP:conf/acl/PapineniRWZ02} can be directly applied to code, many researchers develop code-specific metrics for more reliable evaluation. 
CodeBLEU \cite{DBLP:journals/corr/abs-2009-10297} combines BLEU and code graph match. CrystalBLEU \cite{DBLP:conf/icse/EghbaliP22} excludes common n-grams in a code corpus during BLEU computation to focus on crucial tokens.

% LLM-based metrics score the code with LLM features. BERTScore \cite{DBLP:conf/iclr/ZhangKWWA20} embeds text and the reference answer into representation vectors, before computing their weighted cosine similarity as their semantic similarity. BARTScore \cite{DBLP:conf/nips/YuanNL21} assumes that LLMs are more likely to generate higher-quality text and directly uses the generating probability to evaluate the text. Despite being originally designed for text evaluation, these metrics can be directly adopted to code with minimal changes.

LLM-as-a-judge metrics prompt LLMs to score programs. ICE-Score \cite{DBLP:conf/eacl/Zhuo24} is an early attempt in LLM-based code evaluation with predefined evaluation steps without tool use. 
%CodeJudge \cite{DBLP:conf/emnlp/TongZ24} proposes an analysis and summary pipeline for binary classification on functional correctness, and a JSON report template with manually-curated fault categories for evaluation on requirement adherence. 
CodeVisionary \cite{DBLP:journals/corr/abs-2504-13472} evaluates code with autonomous agents and negotiation. 
MCTS-Judge \cite{DBLP:journals/corr/abs-2502-12468} applies Monte-Carlo Tree Search to predict functional correctness. % SWE-Judge \cite{DBLP:journals/corr/abs-2505-20854} presents different evaluation strategies, and selects the most appropriate strategy with pre-evaluation on a subset.

% However, existing LLM-as-a-judge metrics often produce binary labels to indicate the code's functional correctness, or scores with vaguely-defined criteria such as a 5-point scale on code usefulness. In our work, we reframe the code evaluation task from a more practical and rigorous perspective.

\begin{figure*}[t]
    \centering
    \includegraphics[width=0.93\textwidth]{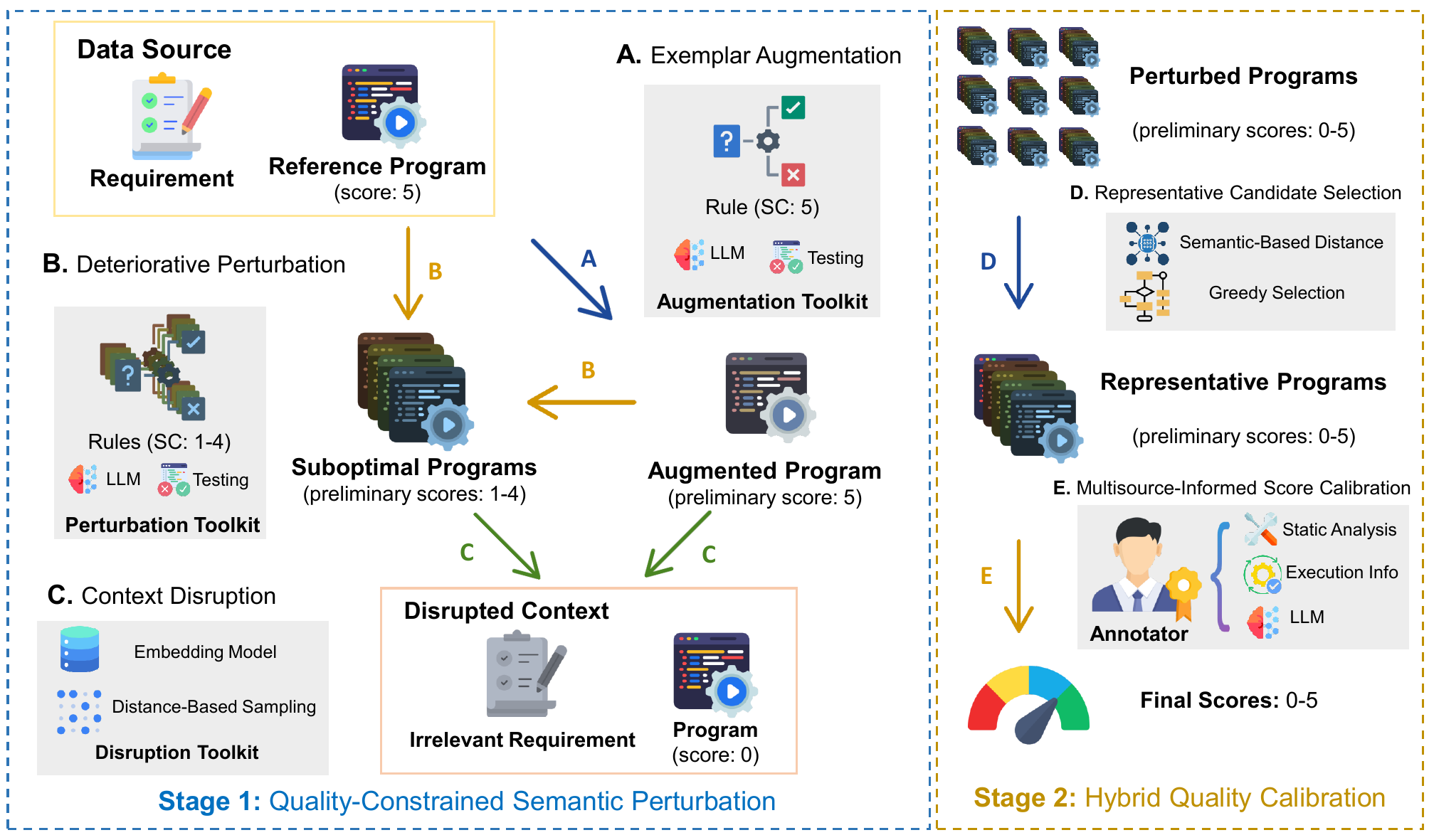}
    \vspace{-0.5em}
    \caption{Overview of \tool. Rule(s) refer to the perturbation rules indicating how the perturbation should be performed, and SC refers to the score ceiling of each rule, as described in Section \ref{sec:pert}.}
    \label{fig:overview}
    \vspace{-1.0em}
\end{figure*}

\subsection{Code Transformation}
% Should we delete or rewrite this subsection? Some studies may appear similar as ours, so we may have to clearly explain why our work is distinguished

Code transformation generates new code based on existing code and transformation rules, having been used to address multiple software engineering tasks, such as code review \cite{DBLP:conf/icse/ThongtanunamPT22} and program repair \cite{DBLP:conf/icse/Li0N20}. 
Recent studies use code transformation to synthesize clone detection benchmarks. %, which require models to predict whether two programs are functionally identical. 
SeqCoBench \cite{DBLP:conf/naacl/MaveliVC25} designs 20 types of single-step Python transformations. EquiBench \cite{DBLP:journals/corr/abs-2502-12466} deploys two-step domain-specific transformations in four domains like CUDA. CETBench \cite{DBLP:journals/corr/abs-2506-04019} performs up to seven transformation steps with seven rule types.

While these transformation workflows can synthesize data for reference-based code evaluation, they are inadequate for our fine-grained evaluation criteria. We contend that modern LLMs are capable enough to perform significantly more intricate and varied transformations. This allows for the development of diverse, general-purpose transformations applicable across various domains.

% \begin{figure}[t]
%     \centering
%     \includegraphics[width=\textwidth]{figures/overview.pdf}
%     \caption{Overview of \tool. Rule(s) refer to the perturbation rules indicating how the perturbation should be performed, and SC refers to the score ceiling of each rule, as described in Section \ref{sec:pert}.}
%     \label{fig:overview}
% \end{figure}

\section{Methodology\label{sec:method}}

% \begin{figure}
%     \centering
%     \includegraphics[width=\textwidth]{figures/rule_demo.pdf}
%     \caption{Examples of perturbation rules.}
%     \label{fig:rule_demo}
%     \vspace{-0.5em}
% \end{figure}

% In this section, we introduce \tool for synthesizing code evaluation data. We formulate the code evaluation task, and then introduce the two stages of \tool. An overview is available in Figure \ref{fig:overview}. We elaborate on Stage 1 in Section \ref{sec:mut} and Stage 2 in Section \ref{sec:calib}.
We present an overview of \tool in Figure \ref{fig:overview}.

\subsection{Code Evaluation: Task Formulation\label{sec:form}}
Given a program \(p\), a metric \(f\) scores the program as \(y=f(p)\). Previous work often used a binary scale \(y\in\{0,1\}\) to indicate \(p\)'s functional correctness, or a 5-point scale with vague qualitative descriptors (e.g., ``somewhat useful'') to assess aspects like usefulness \cite{DBLP:conf/emnlp/TongZ24}. We argue that formulating scores as the effort needed to refine a program until production-ready provides more concrete and objective criteria. 
Thus, we %use an ordinal scale\footnote{An ordinal scale guarantees a natural order, but the intervals between scores are not necessarily equal, unlike interval scales. Thus, the difference between scores of 5 and 4 is not quantitatively the same as the difference between 2 and 1.\label{ft:ord}} to 
measure the refinement effort as our evaluation criteria:

\begin{itemize}
    \item {\bf 5/5:} Production-ready; no effort needed.
    
    \item {\bf 4/5:} Perfect in functionality; requires minor tweaking to enhance code quality.
    
    \item {\bf 3/5:} Perfect in functionality; requires major refactoring to enhance code quality.
    
    \item {\bf 2/5:} Requires minor tweaking to repair functionality.
    
    \item  {\bf 1/5:} Requires major refactoring to repair functionality.
    
    \item {\bf 0/5:} Fundamentally flawed; rewriting is more efficient.
\end{itemize}

Here, a ``minor tweaking'' refers to localized changes that preserve the original control flow and algorithmic structure, like adding a boundary check; a ``major refactoring'' involves altering the control flow or logical structure, like reimplementing an essential data structure. This distinction is inspired by studies on the scales of commits and refactoring \cite{DBLP:journals/dagstuhl-reports/DigGMS14,DBLP:conf/qrs/HonelELW19}.

This scale evaluates the program on two aspects:
\begin{itemize}
    \item {\bf Functionality} measures whether the behavior aligns with the requirement, including input/output, efficiency, etc.
    
    \item {\bf Code quality} measures non-behavioral attributes like readability and maintainability.
\end{itemize}

% We use functionality as the primary criterion. A score of 3 or higher signifies a functionally correct program to be immediately deployed, though it may still require quality improvements.

This precisely defined scale improves upon previous criteria and enables our framework to synthesize data at scale through specifically-designed perturbation and calibration procedures.

% \begin{table}[]
%     \centering
%     \caption{Summary of all perturbation rules.}
%     \begin{tabularx}{\textwidth}{c|X}
%         \toprule

%         \textbf{Score Ceiling} & \multicolumn{1}{c}{\bf Rules Summary} \\ \midrule

%         5 & Reimplementation, alternative algorithm, using standard library containers, functional programming style, code shortening \\ \midrule

%         4 & Vague identifier names, misleading names, wrong case style, confusing operator precedence, magic numbers, long message chains, unused identifier, trivial intermediate variables, dead code, one-liner \\ \midrule

%         3 & Nesting blocks, sequential logic to nesting, replacing flow control statements, massive function inlining, duplicated code, scattered logic, tight coupling, unnecessary abstract structures, global abuse \\ \midrule

%         2 & Off-by-one error, relational operator flipping, logical operator flipping, arithmetic/bitwise operator replacing, incorrect initialization, identifier substitution, argument misorder, wrong enum/constant, unused return value, return removal \\ \midrule

%         1 & Shallow copy, essential step removal, wrong algorithm, loop variable abuse, wrong data structure, wrong recursion base, object lifecycle mismanagement, inplace modification, validation logic removal, serialized data misinterpret \\ \bottomrule
%     \end{tabularx}
%     \label{tab:rule_sum}
%     \vspace{-0.5em}
% \end{table}

\begin{table}[t]
    \centering
    \small
    \caption{Summary of Perturbation Rules by score ceiling \(c(u)\). The rule set contains 45 rules total.}
    \vspace{-0.5em}
    \label{tab:rule_sum}
    \renewcommand{\arraystretch}{1.1}
    \begin{tabularx}{\columnwidth}{c l c >{\raggedright\arraybackslash}X}
        \toprule
        $c(u)$ & \textbf{Category} & \textbf{\#} & \textbf{Examples (\(\mathcal{I}_u\))} \\
        \midrule

        5 & Alternatives & 5 &
        \textbullet~Refactor to functional style (e.g., streams, lambdas). \newline
        \textbullet~Replace primitive structures with standard library equivalents. \newline
        \textbullet~Shorten code using ternary operators or comprehensions. \\
        \midrule

        4 & Readability & 10 &
        \textbullet~Replace descriptive names with generic single-letter variables. \newline
        \textbullet~Introduce unreachable code. \newline
        \textbullet~Introduce magic numbers. \\
        \midrule

        3 & Design & 9 &
        \textbullet~Create ``Long Methods'' by inlining helper functions. \newline
        \textbullet~Introduce duplicated code by inlining function calls. \newline
        \textbullet~Increase nesting depth of conditional logic. \\
        \midrule

        2 & Minor Bugs & 11 &
        \textbullet~Introduce off-by-one errors. \newline
        \textbullet~Swap binary operators (e.g., $a+b \rightarrow a-b$). \newline
        \textbullet~Replace deep copies with shallow ones. \\
        \midrule

        1 & Major Defects & 10 &
        \textbullet~Remove recursion base cases. \newline
        \textbullet~Replace algorithms with plausible but wrong ones. \newline
        \textbullet~Remove essential input validation logic. \\
        \bottomrule
    \end{tabularx}
    \vspace{-1.0em}
\end{table}

\subsection{\STAGEa\label{sec:mut}}

Given a code generation benchmark, where each sample is a triplet \((r,p_\text{ref},t)\), where \(r\) is the requirement, \(p_\text{ref}\) is the reference program, and \(t\) is the set of unit tests, this stage creates a code evaluation benchmark of \((r,p,s(p))\) where \(p\) is a program whose target score is \(s(p)\) under our evaluation criteria. Previous work \cite{DBLP:journals/pacmse/WangGGFCX25,DBLP:journals/corr/abs-2504-13472} often directly prompts LLMs to generate programs for requirement \(r\), leading to uncontrolled error types and unbalanced score distributions. % For example, the human-annotated scores from \cite{DBLP:journals/pacmse/WangGGFCX25} are heavily skewed towards higher scores. 
To address these issues, we propose a novel perturbation stage to synthesize program \(p\) and compute score \(s(p)\).

In this section, we first introduce our perturbation design, and then describe steps A, B, C in Figure \ref{fig:overview}:

\begin{itemize}
    \item \textbf{\Stageaa}: generates more 5-point programs besides existing reference programs.
    \item \textbf{\Stageab}: synthesizes 1- to 4-point programs from 5-point ones.
    \item \textbf{\Stageac}: produces 0-point programs.
\end{itemize}

\subsubsection{Perturbation Rules\label{sec:pert}}
% In this subsection, we introduce the perturbation process, which helps create a diverse set of requirement-program-quality triples, enhancing data variability and quantity, which is essential for comprehensively evaluating LLM-as-a-judge metrics.
To comprehensively evaluate LLM-as-a-judge metrics, we utilize perturbation, which transforms programs across the evaluation scale defined in Section \ref{sec:form}, creating a diverse set of \textit{requirement-program-quality} triples.
% A perturbation step \(p^\prime=f_u(p)\) transforms program \(p\) into \(p^\prime\) with perturbation rule \(u\). \(u\) consists of a natural language instruction on how to perform the code transformation, and a score ceiling \(c(u)\in\{1,2,3,4,5\}\), indicating that the perturbed program's score \(s(p^\prime)\) is capped at \(c(u)\). 
Formally, a perturbation $p^\prime = f_u(p)$ transforms a program $p$ given a perturbation rule $u$. Each rule $u$ is a tuple $\langle \mathcal{I}_u, c(u) \rangle$. $\mathcal{I}_u$ is a natural language instruction describing the transformation, and $c(u) \in \{1, \dots, 5\}$ is the \textbf{score ceiling},
% The score ceiling is designed to match our evaluation criteria. For instance, \(c(u)=5\) means that the perturbation reimplements the program with negligible impact on its functionality and quality, while \(c(u)=1\) indicates the introduction of a functionality fault requiring major refactoring to repair. Therefore, given the original program \(p\) and its score \(s(p)\), after a perturbation sequence \(u_1,...,u_n\), the score of the resulting program \(p^\prime=f_{u_n}(...f_{u_1}(p))\) equals to \(\min(s(p),c(u_1),...,c(u_n))\).
capping the maximum score $p^\prime$ can achieve, i.e., acting as bottlenecks to the program quality. The post-perturbation score is $s(p^\prime) = \min(s(p), c(u))$. For instance, applying a rule $u$ where $c(u)=2$ to a program of $s(p)=5$ degrades it to 2 points, now requiring minor functionality repairs.

% \textbf{Perturbation rule creation.} Each rule is designed with the following principle:

% \begin{itemize}
%     \item {\bf Generalizability.} All rules should be generalizable across programming languages and domains, instead of relying on certain language or library features.

%     \item {\bf Authenticity.} Rules with score ceilings \(c(u)<5\) should mimic common real-world errors.

%     \item {\bf Controllability.} Each rule defines a specific transformation mapping to a specific score indicated by its score ceiling, allowing us to precisely control the resulting program's score.
% \end{itemize}

\textbf{Rule Creation.}
% Inspired by previous work \cite{DBLP:conf/emnlp/XuWPSFWL23}, which manually curates few-shot samples and prompts LLMs to generate more, we adopt the following rule creation pipeline for each target score \(s(p)\):
We curate 45 rules with these steps:

\begin{enumerate}
    % \item We manually write three rules from previous work, like \cite{DBLP:conf/naacl/MaveliVC25}, as seeds.
    \item \textbf{Seed Selection:} Manually drafts three rules reflecting common code defects and smells for each score ceiling.
    % \item We prompt an LLM with these seeds and our scoring criteria to generate candidate rules.
    \item \textbf{LLM Expansion:} Prompts the state-of-the-art Gemini 2.5 Pro to generate more rules given the seed rules, strictly adhering to the definitions of ``minor tweaking'' and ``major refactoring'' outlined previously.
    % \item We manually remove candidate rules whose instructions fail to match their score ceilings, and prompt the LLM again until no more unique rules can be generated.
    \item \textbf{Validation:} Manually verifies that each instruction $\mathcal{I}_u$ matches its score ceiling $c(u)$.
\end{enumerate}

\textbf{Rule Taxonomy.}
% We use Gemini 2.5 Pro \cite{DBLP:journals/corr/abs-2507-06261}, a state-of-the-art general-purpose LLM, for rule creation, resulting in 45 rules in total. Examples of perturbation rules with each possible score ceiling are shown in Figure \ref{fig:rule_demo}. All remaining rules can be found in our repository, and we summarize all rules in Table \ref{tab:rule_sum}.
Rules are categorized by their score ceiling, mapping directly to the specific degradation required to reach that score level, as summarized in Table \ref{tab:rule_sum}.

\begin{itemize}
    \item \textbf{(Level 5) Semantic-Preserving Variations:} generates alternative implementations that maintain functionality and high code quality, involving stylistic adaptations.
    
    \item \textbf{(Level 4) Minor Code Quality Degradation:} introduces localized readability or style issues while preserving perfect functionality. Addressing these issues requires only ``minor tweaking'' as the control flow remains intact.
    
    \item \textbf{(Level 3) Major Code Quality Degradation:} introduces design flaws that do not affect functionality, requiring ``major refactoring'' altering the program structure.
    
    \item \textbf{(Level 2) Minor Functional Defects:} injects specific, localized bugs that break functionality, but can be repaired with ``minor tweaking'' without rewriting the core logic.
    
    \item \textbf{(Level 1) Major Functional Defects:} renders the program functionally flawed, requiring ``major refactoring'' of structural overhauls, but not a complete rewrite.
\end{itemize}

% For perturbation, due to modern LLMs' powerful code editing capabilities, we prompt an LLM to perform perturbation. The LLM is instructed to first analyze the original program \(p\), and determine whether the given rule \(u\) is feasible. If not, the LLM refuses to perturb. Otherwise, it plans the changes and outputs the perturbed program \(p^\prime\). We demonstrate the perturbation process in Algorithm \ref{algo:mut}, which performs at most \(N_{\max}\) perturbation steps, each with a perturbation rule. In the algorithm, we first randomly select the number of perturbation steps \(N\) and sample score ceilings \(c_1...c_N\) of all rules given the target score \(s(p^\prime)\). During step \(i\), we sample a rule with score ceiling \(c_i\) and invoke the LLM to perturb. We return the final program if all steps succeed.
\textbf{Perturbation Algorithm.}
As shown in Algorithm \ref{algo:mut}, the process uses an LLM for multi-step degradation. Given a target score $s(p^\prime)$, we sample a sequence of rules $u_1, \dots, u_N$ such that $\min_ic(u_i) = s(p^\prime)$. The LLM first validates if a rule is applicable to the specific program $p$ (e.g., verifying recursions exist before removing its base cases) and then generates the perturbed program $p^\prime$. We check its functionality via unit tests.

\begin{algorithm}[t]
\caption{Multi-Step Perturbation} \label{algo:mut}

\renewcommand{\algorithmicrequire}{\textbf{Input:}}
\renewcommand{\algorithmicensure}{\textbf{Output:}}

\renewcommand{\arraystretch}{0.5}
\begin{algorithmic}
    \REQUIRE Program $p$, Perturbation rule set $U_*$, Target score $s(p^\prime)$
    \ENSURE Perturbed program $p^\prime$

    \STATE \textit{// 1. Generate score ceilings}
    \STATE $N \gets \text{random}(1,N_{\max})\textit{ // Number of perturbations}$ 

    \FOR{$i=1$ to $N$}
        \STATE $c_i=\text{random}(s(p^\prime),5)\textit{ // Score ceiling of step }i$ 
    \ENDFOR

    \STATE \textit{// Ensure }$\min_i c_i=s(p^\prime)$
    \STATE $j \gets \text{random}(1,N)$ 
    \STATE $c_j=s(p^\prime)$ 

    \STATE
    \STATE \textit{// 2. Follow randomly-selected rules}
    \STATE $p_0\gets p$
    \FOR{$i=1$ to $N$}
        \STATE $u\gets \text{sample}(U_{c_i}) \textit{ // }U_{c_i}\textit{: rules with score ceiling }c_i$ 

        \IF{\NOT $\textit{LLM\_feasible}(p_{i-1},u)$}
            \RETURN$\textsc{null}\textit{ // Fails if }u\textit{ is infeasible on }p_{i-1}$
        \ENDIF
        \STATE $p_i\gets \textit{LLM\_perturb}(p_{i-1},u)\textit{ // Performs perturbation}$ 
    \ENDFOR
    \STATE $p^\prime\gets p_N$
    \RETURN $p^\prime$
        
\end{algorithmic}
\renewcommand{\arraystretch}{0.1}
\end{algorithm}

\subsubsection{\STAGEaa}
Code generation benchmarks typically provide a single reference program per requirement. Relying on this solitary seed would constrain perturbation diversity, limiting benchmark comprehensiveness. Thus, we enrich the initial seed pool with additional 5-point programs to ensure a diverse foundation for future perturbation. 

% The process begins with a pre-perturbation check to verify the initial reference programs with unit tests. Failing programs due to incorrect outputs or environmental incompatibilities are discarded. We only keep requirements with at least one verified reference program for subsequent perturbation.

For each requirement, we keep existing reference programs as the initial seed pool of 5-point programs, and add new programs until the pool has \(M\) programs. 
To generate a new program \(p^\prime\), we invoke Algorithm \ref{algo:mut} with target score \(s(p^\prime)=5\) and a random existing program \(p\). The algorithm utilizes perturbation rule set \(U_5\) (rules with score ceiling 5), which rewrites \(p\) while preserving its functionality. 
% The rules in \(U_5\)  produce significant structural changes, such as shifting programming paradigms (e.g., to functional programming) or implementing an alternative algorithm. Therefore, one perturbation step (\(N_{\max}=1\)) is considered sufficient. 
The resulting program \(p^\prime\) is verified via unit testing.

\subsubsection{\STAGEab}
This step generates suboptimal programs of 1 to 4 points. Previous code transformation work \cite{DBLP:conf/naacl/MaveliVC25} often applied a single, minor change like altering an operator, resulting in benchmarks with low diversity and overly localized flaws. In contrast, our approach features richer defect categories of both localized and structural changes to better reflect complex real-world coding errors.
 
To generate a program \(p^\prime\) with target score \(s(p^\prime)\in\{1,2,3,4\}\), 
% we begin by randomly selecting a 5-point program \(p\). We then apply Algorithm \ref{algo:mut}, which executes a sequence of up to \(N_{\max}=5\) perturbation steps to \(p\). 
we apply Algorithm \ref{algo:mut} to a random 5-point program \(p\) with a sequence of perturbations to obtain \(p^\prime\).
% This multi-step process is designed to create programs with more complex and realistic flaws than a single change.
% For each successfully generated program, we confirm that its functional status aligns with its target score via unit testing: programs targeting scores of 3 and 4 must pass all tests, while those targeting scores of 1 or 2 must fail at least one. Any programs failing to meet this condition are discarded. This generation and validation cycle is repeated until we have collected \(M\) programs for each target score per requirement.
We run unit tests on \(p^\prime\), and keep \(p^\prime\) only if it passes all tests (\(s(p^\prime)\in\{3,4\}\)) or fails at least one test (\(s(p^\prime)\in\{1,2\}\)). This generation--validation cycle is repeated until \(M\) programs are collected per target score per requirement.

% A key advantage of this approach is its efficiency. We are concerned only with the final functional status of the generated program---whether it passes or fails tests as required by its target score. This outcome-based approach eliminates the expensive need to verify each intermediate perturbation step. While each program's functional status is confirmed at this stage, we leave the assessment of its code quality and the determination of its final, calibrated score to Section \ref{sec:calib}.
This outcome-based design is efficient: Verifying each intermediate perturbation is unnecessary; we only require the functional status of the final \(p^\prime\) matches \(s(p^\prime)\). We defer code quality evaluation and score calibration to Section \ref{sec:calib}.

\subsubsection{\STAGEac}
This step creates 0-point programs. Instead of directly generating buggy code, we pair programs with semantically unrelated requirements. This approach creates reliable test cases that assess LLMs' ability to detect fundamental contextual irrelevance. Furthermore, since these pairings are incorrect by definition, their scores of 0 are absolute, bypassing manual calibration.

We use an embedding model \(E\) to embed requirements \(r_1,...,r_n\) into vectors \(E(r_1),...,E(r_n)\), and then compute a distance matrix \(D\) with cosine distance: \(D_{ij}=1-\frac{E(r_i)\cdot E(r_j)}{\|E(r_i)\|\|E(r_j)\|}\in [0,2]\), where higher values indicate greater semantic dissimilarity.

For a given requirement \(r_i\), we sample \(M\) distinct requirements by applying softmax over \(D\) with temperature \(\tau=0.03\). Requirement \(r_j\)'s probability of being chosen is \(P_{ij}=\frac{\exp (D_{ij}/\tau)}{\sum_{k=1}^n\exp (D_{ik}/\tau)}\).
This encourages conceptually distant requirements, creating unambiguous 0-point pairings.

Finally, for each sampled requirement, we randomly select one of its associated programs to serve as a 0-point program for requirement \(r_i\).

\subsection{\STAGEb\label{sec:calib}}
The preceding stage yields a large corpus of programs, where each program \(p\) is associated with a requirement \(r\) and a target score \(s(p)\). While its functional status is confirmed to align with \(s(p)\), its non-behavioral code quality and refinement effort remain unexamined. 
To determine the final score \(s\), the conventional annotation procedure would require human experts to assess both the functionality and code quality of each program,
% a process that imposes a significant cognitive load and demands deep domain expertise, making it prohibitively slow and often unreliable.
which is often slow and unreliable due to the immense cognitive load and domain knowledge required.

To address these challenges, we design the \stageb stage for both efficiency and reliability. Its core principle is to decouple the assessment of functionality and non-behavioral code quality. As the functionality has already been examined in the previous stage, annotators no longer need to reason through complex program logic. Instead, their task is narrowed to a much simpler and more objective judgment: reviewing the \texttt{diff} between program \(p\) and its 5-point pre-perturbation parent, determining its code quality and scope of refinement changes with multisource information, and labeling the final score \(s\).

This decoupled approach dramatically reduces cognitive load on annotators, accelerating annotation and increasing scoring reliability. This efficiency allows us to curate a much larger verified benchmark than a few hundred samples \cite{DBLP:journals/pacmse/WangGGFCX25} while keeping our fine-grained evaluation criteria.

This stage contains two steps (D and E in Figure \ref{fig:overview}). We select a representative subset of programs, and then perform the final, streamlined manual annotation to calibrate the scores.

\subsubsection{\STAGEba}
The goal of this step is to select a balanced and diverse subset. For each target score, the previous stage yields a large pool of \(N=nM\) programs, where \(n\) is the number of requirements, and \(M\) is the number of programs per target score. From this pool, we aim at curating a representative subset of \(m\) programs per target score before calibration.

To ensure diversity and avoid redundancy, we quantify the dissimilarity between programs with normalized token edit distance, which is more semantically aware than character-wise metrics. We first tokenize each program, discarding comments and whitespace, and then compute distance matrix \(L\). \(L_{ij}\) is the minimum number of token edits (insertions, deletions, and substitutions) to transform program \(p_i\) into \(p_j\), normalized by the token count of the longer program.

With the distance matrix \(L\), we then select \(m\) candidates from the \(N\) programs with the same target score with a greedy approach, which employs a farthest-point heuristic. It begins with a random program, and iteratively adds the remaining \(m-1\) candidates. In each iteration, we aim to select the most dissimilar program to the candidate set \(A\). To achieve this, we compute the minimum distance from each program to any candidates, and then greedily choose an unselected program to maximize this distance. We repeat this process until \(m\) programs are selected.

% \begin{algorithm}[t]
% \caption{Candidate Selection} \label{algo:select}

% \renewcommand{\algorithmicrequire}{\textbf{Input:}}
% \renewcommand{\algorithmicensure}{\textbf{Output:}}
% \renewcommand{\arraystretch}{0.1}
% \begin{algorithmic}
%     \REQUIRE Distance matrix \(L\), Program and candidate count \(N,m\)
%     \ENSURE Candidate set \(A\)

%     \STATE \texttt{// Begin with a random candidate}
%     \STATE $A=\{\text{random}(1,N)\}$
%     \STATE \texttt{// Iteratively select the remaining }$m-1$\texttt{ candidates}
%     \FOR{$i=2$ to $m$}
%         \FOR{$j=1$ to $N$}
%             \STATE \texttt{// }\texttt{min distance from program }$j$\texttt{ to candidates}
%             \STATE $d_j=\min_{k\in A}L_{jk}$
%         \ENDFOR
%         \STATE \texttt{// Maximize ``min distance'' among the unselected}
%         \STATE $A\gets A\cup \{\text{argmax}_{j\not\in A}d_j\}$
%     \ENDFOR
%     \RETURN $A$
% \end{algorithmic}
% \renewcommand{\arraystretch}{0.1}
% \end{algorithm}

\subsubsection{\STAGEbb}
This final step calibrates the target score \(s(p)\) of each candidate program \(p\) into the final, human-verified score \(s\). To facilitate an evidence-informed calibration, the annotator is provided with a multisource report:

\begin{itemize}
    \item The \texttt{diff} comparison between \(p\) and its 5-point parent.
    % \item The target score \(s(p)\).
    \item The sequence of perturbation rules.
    \item A static analysis report.
    \item The full execution and test report.
    \item An LLM-generated code quality report.
\end{itemize}

As the functional status of each program has already been confirmed via automated testing, the annotator's judgment is focused entirely on the code quality and the scope of changes in the \texttt{diff}. Specifically, for programs with functional flaws, the annotator verifies whether the functional changes in the \texttt{diff} are structural refactors for \(s=1\), or a localized tweak for \(s=2\); for functionally correct programs, the annotator first assesses whether the code quality is perfect\footnote{The outputs from static analysis tools and the LLM are strictly advisory. The final judgment on code quality depends solely on the human annotator.}, and similarly judges the scope of changes to distinguish programs with scores of 3--5. 

After receiving the final score \(s\), we obtain the final benchmark of requirement \(r\), program \(p\), and ground-truth score \(s\).
\section{Benchmark and Study Design\label{sec:setup}}
This section describes \bench, a new code evaluation benchmark created by \tool. It then outlines the design of our empirical study, which leverages \bench to evaluate recent LLM-as-a-judge metrics for code.

\subsection{\bench: Curation}
We use four recent, challenging benchmarks as data sources:

\begin{itemize}
    \item {\bf BigCodeBench} \cite{DBLP:conf/iclr/ZhuoVCH0WYZHPB025} focuses on diverse Python libraries and complex instructions. We use the {\bf Hard} subset. 
    % is a Python benchmark emphasizing the LLM capability of using diverse API invocations as tools, and understanding complex instructions. It features 1140 coding tasks with APIs from 139 libraries, each accompanied by tests with high branch coverage. We use the {\bf Hard} subset, featuring 148 difficult tasks with multiple API calls and longer reference programs.
    \item {\bf LiveCodeBench} \cite{DBLP:conf/iclr/JainHGLYZWSSS25} requires LLMs to reason on difficult algorithms. We use tasks from \textbf{v3--v6} with {\bf Hard} difficulty in C++, due to its dominance in competitive programming\footnote{Most submissions at https://codeforces.com/problemset/status are in C++.}.
    % features competitive programming problems from CodeForces, AtCoder, and LeetCode, challenging LLMs' capability to reason about complicated algorithms. The authors provide continuous updates with newly released problems to mitigate data contamination. We use problems from versions {\bf v3--v6} utilizing standard input/output and with {\bf Hard} difficulty. For these problems, we use C++ as the target language to increase language diversity and reflect its dominance in competitive programming\footnote{For instance, most recent submissions at https://codeforces.com/problemset/status are in C++.}.
    \item {\bf APPS} \cite{DBLP:conf/nips/HendrycksBKMAGB21} contains 5,000 comprehensive Python tasks from various platforms. We use {\bf competition} problems.
    % is a Python benchmark with a comprehensive test set of 5000 problems of various difficulty, from platforms including Kattis and Codeforces. Each problem comes with a natural language description and high-quality test cases. We select {\bf competition-level} problems that come with reference programs, leaving us with 310 problems.
    \item {\bf Aider-Polyglot}\footnote{https://aider.chat/2024/12/21/polyglot.html} features difficult, multilingual tasks to test LLMs' cross-file editing and algorithm design capability. We use problems in Java, Python, and C++.
    % is a multilingual benchmark based on 225 difficult problems from Exercism, challenging the LLMs' reasoning and code editing capabilities. While Aider-polyglot is a code editing benchmark, we adapt it for code generation by extracting the problem statements and treating each as a challenge to implement from scratch. We select the 107 problems configured for Java, Python, and C++.
\end{itemize}

% To curate a diverse code evaluation benchmark, 
We begin by randomly sampling up to 100 coding tasks from each selected benchmark as requirements \(r\). Their reference programs \(p_\text{ref}\) serve as the initial, 5-point programs for our benchmark after being verified on test cases. % Since LiveCodeBench does not contain reference programs, we generate one per requirement using GPT-5 and Claude-Sonnet-4\footnote{We use GPT-5-20250807 with minimal reasoning effort and Claude-Sonnet-4-20250514 without reasoning for all experiments.}, and verify the generated program via testing.

As the \stagea stage requires numerous LLM invocations, we sample one of these cost-efficient LLMs without reasoning for each perturbation step: DeepSeek-V3.1, Qwen3-Coder-480B, Kimi-K2, and Gemini-2.5-Flash. We generate \(M=2\) programs per target score per requirement with temperature \(T=0.3\) and an output token limit of 8192. For the embedding model, we use OpenAI's text-embedding-3-large. All LLM invocations are made online via API providers. Unit tests are executed locally on an Ubuntu server with two Intel Xeon Platinum 8276 CPUs and 256 GB RAM, in Bubblewrap's sandbox environment, with constraints of 5s CPU time and 12 GB RAM. % restricting write access outside working and temporary directories. % If a requirement's reference program \(p_\text{ref}\) attempts to bypass this restriction, the requirement and all associated programs are discarded.

% \begin{table}[]
%     \centering
%     \caption{LLMs used for perturbation. Prices are collected from official documentation without considering context caching or third-party API providers.}
%     \label{tab:mutllm}
%     \begin{tabular}{cc|cc|cc|c}
%         \toprule
%         \multirow{2}{*}{\bf LLM} & \multirow{2}{*}{\bf Release Date} & \multirow{2}{*}{\bf Size} & \multirow{2}{*}{\bf Reasoning} & \multicolumn{2}{c|}{\bf Price/M tokens} & \multirow{2}{*}{\bf Developer}  \\
%         &&&& Input & Output \\
%         \midrule
%         DeepSeek-V3.1 & 2025.8 & 671B & Disabled & \$0.56 & \$1.68 & DeepSeek AI \\
%         Qwen3-Coder-Instruct & 2025.7 & 480B & No & \$1.50 & \$7.50 & Alibaba \\
%         Kimi-K2-Preview & 2025.7 & 1T & No & \$0.60 & \$2.50 & Moonshot AI  \\    
%         Gemini-2.5-Flash & 2025.6 & - & Disabled & \$0.30 & \$1.00 & Google \\
%         \bottomrule
%     \end{tabular}
% \end{table}

In the \stageb stage, we tokenize programs with Pygments lexers.
%, and discard programs with less than 200 non-whitespace tokens or 10 lines of code 
We select up to \(m=90\) programs per data source per target score. We use DeepSeek-V3.1 for LLM-generated reports, and cppcheck (C++), PMD (Java), and Pylint (Python) for static analysis. We manually calibrate scores of \textbf{all} programs, including reference programs \(p_\text{ref}\).

\subsection{Annotation Details and Statistics\label{sec:prelim}}
The manual calibration process is performed by a primary expert with 14 years of programming experience. After the 14-hour annotation procedure (25 seconds/program), 1,431 programs (73\% of 1,962) retain their target scores without modification, indicating that target scores are already highly accurate. 
To validate the objectivity and reproducibility of our scoring criteria, we conduct an annotation study with 10\% sampled data (196 programs) from \bench, and calibrate the scores with two experts, each with more than five years of programming experience. The inter-rater agreement results show high agreement with a Cohen's \(\kappa\) of 96.7\%, indicating that our criteria are highly objective and reproducible. This justifies our choice of a primary annotator to optimize annotation cost without sacrificing benchmark reliability.
% a Krippendorff's \(\alpha\) of 96.40\%, an ICC(2, 1) of 96.47\%, and an Exact Match score\footnote{The percentage of programs where all experts assign the same final score.} of 76.02\%. 

The final \textbf{\bench-Full} has an average of 58.4 lines of code, 8.5 cyclomatic complexity, and 477 tokens per program.
We compare the statistics of \bench with other benchmarks in Table \ref{tab:benchstats}. Specifically, we report \(\kappa\) for human annotation reliability, compare different evaluation criteria, and report Normalized Shannon Entropy \(S=-\sum_{k=1}^Np_k\log_Np_k\in[0,1]\) to indicate uniformity, where \(p_i\) is the frequency of the \(i\)-th possible score, and \(N\) is the number of possible scores. \(S=1\) indicates a perfectly uniform quality distribution, whereas lower values indicate class imbalance. As Table \ref{tab:benchstats} demonstrates, \bench achieves a very high entropy of \(S=0.995\), indicating a remarkably more balanced score distribution than other benchmarks.
    
\begin{table}[]
    \centering
    \caption{Benchmark statistics.  \textbf{Rectifiable} refers to \textbf{\bench} without 0-point programs, used in RQ1. \textbf{Full} refers to the complete \textbf{\bench}, used in other experiments. \textbf{\#Prog.}/\textbf{\#Req.} means the number of programs/requirements. \(\kappa\) on our benchmark is computed on 10\% data (196 programs). Execution-based benchmarks (EX marked as \Checkmark) only evaluate code on functionality, and due to their absence of manual annotation, \(\kappa\) is inapplicable to them. }
    % \textbf{Unverified} refers to \textbf{\bench-Unverified} (before \stageb).
    \begin{tabular}{cc|cc|cc}
        \toprule
        
        {\bf Benchmark} & {\bf EX} & {\bf \#Prog.} & {\bf \#Req.} & \(\kappa\) & \(S\) % & {\bf LoC} & {\bf CC} & {\bf \#Tokens} 
        \\ \midrule

        % \textbf{Unverified} & 1,158 & 909 & 1,038 & 942 & 1,045 & 2,597 & 7,689 & 40.1 & 6.3 & 330 & 398 \\
        % \textbf{Full} & 330 & 316 & 335 & 363 & 361 & 257 & 1,962 & 58.4 & 8.5 & 477 & 388 \\ 
        % \textbf{Rectifiable} & 0 & 316 & 335 & 363 & 361 & 257 & 1,632 & 59.1 & 8.8 & 482 & 373 \\ \bottomrule

        \textsc{\bench-Full} & \multirow{2}{*}{\XSolidBrush} & 1,962 & 388 & \multirow{2}{*}{96.7} & 0.997 % & 58.4 & 8.5 & 477 & 
        % & {0--5, Repair Effort} 
        \\
        \textsc{\bench-Rectifiable} & & 1,632 & 373 & & 0.995 % & 59.1 & 8.8 & 482 
        % & {1--5, Repair Effort}
        \\ \midrule

        Wang et al. \cite{DBLP:journals/pacmse/WangGGFCX25} & \XSolidBrush & 300 & 100 & 82.2 & 0.950 % & 1--5, Subjective Likert 
        \\
        CodeVisionary \cite{DBLP:journals/corr/abs-2504-13472} & \XSolidBrush & 363 & 121 & 80.0 & 0.675 % & 0--4, Subjective Likert 
        \\ 
        % CoNaLa \cite{DBLP:journals/jss/EvtikhievBSB23} & \XSolidBrush & 2,360 & 472 & - &  0.901 \\ 
        CodeJudge-Eval \cite{DBLP:conf/coling/ZhaoLT0YL025} & \Checkmark & 1,860 & 457 & - & 0.785 % & Nominal, Execution Status 
        \\ 
        CodeJudgeBench \cite{DBLP:journals/corr/abs-2507-10535} & \Checkmark & 10,704 & 5,352 & - & 1.000 % & Binary, Pass/Fail  
        \\ 
        \bottomrule
          
    \end{tabular}
    \label{tab:benchstats}
    \vspace{-1.0em}
\end{table}

\subsection{Research Questions}
We study these research questions with \bench:

\begin{itemize}
    \item {\bf RQ1: How do existing code evaluation metrics perform on \bench?}

    % \footnote{Pearson's \(r\) is not used, since it measures linear relationships and requires interval scales. Applying to our ordinal-scaled scores would incorrectly assume identical gaps between consecutive scores, as explained in Footnote \ref{ft:ord}.}
    This question assesses the performance of existing metrics by measuring the alignment between scores from each metric and ground-truth scores. To achieve this, we use two correlation coefficients, Spearman's \(\rho\) and Kendall's \(\tau\), which measure their ranked relationship, as well as Krippendorff's \(\alpha\) % and Quadratic Weighted \(\kappa\), measuring 
    to measure their absolute agreement. We also measure the mean score of each metric to study systematic biases, and the Earth Mover's Distance (EMD) \cite{DBLP:conf/iccv/RubnerTG98} to study score distribution alignment.
    
    \item {\bf RQ2: How does metric performance vary across subtasks of various complexity?} 

    This question provides a fine-grained diagnosis of metric performance by isolating evaluation subtasks: refinability gating, code quality assessment, functionality verification, and refinement effort estimation. Each can be formulated as a binary classification task. We use Matthews Correlation Coefficient (MCC) for its symmetric assessment of both positive and negative classes. Unlike \(F_1\)-score and accuracy, MCC effectively penalizes random guessing.
    
    \item {\bf RQ3: Can these metrics evaluate code consistently?}

    A reliable evaluation metric must assign consistent scores to the same program when invoked multiple times. We run each metric five times on the same program on the Aider-Polyglot subset, before quantifying consistency with two agreement coefficients, namely Krippendorff's \(\alpha\) and intra-class correlation (specifically, ICC(3, 1)).
    
\end{itemize}

% WIP: due to resource limitations, we only use all samples for RQ1-2, and 500 samples for RQ3.

\subsection{Experimental Setup}
We study several code metrics with \bench:

\subsubsection{Rule-Based Metrics} These metrics compute rule-based similarity between program \(p\) and the reference program \(p_\text{ref}\).

\begin{itemize}
    \item {\bf ChrF++} \cite{DBLP:conf/wmt/Popovic17} computes \(F_2\)-scores based on character and token n-grams. While less commonly used than BLEU, it performs better on code evaluation \cite{DBLP:journals/pacmse/WangGGFCX25}.
    \item {\bf CodeBLEU} \cite{DBLP:journals/corr/abs-2009-10297}, tailored for code evaluation, combines BLEU, weighted n-gram match, syntactic AST match, and semantic data-flow match into the final score.
    \item {\bf Edit Similarity} 
    \cite{levenshtein1965binary} is derived from the edit distance. If the token edit distance between \(p\) and \(p_\text{ref}\) is \(d(p,p_\text{ref})\), their edit similarity is \(1-\frac{d(p,p_\text{ref})}{\max(|p|,|p_\text{ref}|)}\).
\end{itemize}

To standardize these metrics for a fair comparison, we first tokenize each program \(p\) with Pygments lexers instead of general-purpose LLM tokenizers to better capture the syntactic code structure. After discarding whitespace and comments, we compute raw score \(s_p\) with the metric. As these metrics produce continuous scores on varying scales, we normalize them into our 0--5 integer scale by \(round(5(s_p-s_\text{min})/(s_\text{max}-s_\text{min}))\), where \(s_{\min}\) and \(s_{\max}\) are the minimum and maximum raw scores from that metric.

% \subsubsection{LLM-Feature Based Metrics}

% \begin{itemize}
%     \item \textbf{CodeBERTScore}
%     \item \textbf{GPTScore}
% \end{itemize}

% {\bf\color{cyan}--- WIP: min-max normalization \& rounding ---}

% {\bf\color{orange}--- If we don't have enough time for experiments, we might as well delete this category; they are seldomly used anyways ---}

\subsubsection{LLM-as-a-Judge Metrics} These metrics prompt the LLM to respond with a score, sometimes using a predefined workflow to collect additional context before scoring.

\begin{itemize}
    \item {\bf ICE-Score} \cite{DBLP:conf/eacl/Zhuo24} prompts the LLM to follow several predefined evaluation steps and assign the final score. % on code usefulness and functional correctness.
    \item {\bf CodeJudge} %\footnote{The original work also designs another ``analyze-then-summarize'' pipeline for binary classification on functional correctness. We do not use this pipeline as our work focuses on a more fine-grained evaluation.} 
    \cite{DBLP:conf/emnlp/TongZ24} first prompts an LLM to detect faults and computes a rule-based score with the JSON fault report. As the score computation misaligns with our evaluation criteria, we instead prompt the LLM to generate the final score based on our criteria with the report.
    \item {\bf CodeVisionary} \cite{DBLP:journals/corr/abs-2504-13472} gathers context with an agentic tool-use workflow, and then lets several LLM judges discuss and score the code. Due to resource constraints, the maximum number of tool invocations is reduced to 20. % We disable the screenshot and interaction actions due to the absence of front-end development in our benchmark.
\end{itemize}

We use DeepSeek-V3.1 and Gemini-2.5-Flash for LLM-as-a-judge metrics. We adhere to these metrics' original prompt templates and hyper-parameter configurations, but replace their evaluation criteria with ours. For the ease of reading, we multiply the coefficients by 100.

\section{Study Results\label{sec:res}}

\subsection{RQ1: Performance and Score Distribution}
\begin{table}[]
    \centering
    \caption{Performance of each metric. The best result in each column is underlined and marked bold. Results at most 3\% worse than it are underlined.}
    \begin{tabular}{c|ccc|cc}
        \toprule

        {\bf Metric} & \textbf{\(\rho\)} & \textbf{\(\tau\)} & \textbf{\(\alpha\)} & \textbf{EMD} & \textbf{Mean} \\ \midrule

        \multicolumn{6}{c}{\vspace{0.2em} \textbf{\textsc{Part I: \bench-Full}} \vspace{0.2em}} \\ \midrule

        \rowcolor{gray!25}\multicolumn{6}{c}{\bf Rule-Based Metrics} \\ \midrule
        
        ChrF++          &  46.18 &  37.48 &  44.57 & \underline{\textbf{0.308}} & 2.703 \\
        CodeBLEU        &  43.92 &  35.77 &  42.12 & \underline{0.326} & 2.477 \\
        Edit Similarity &  43.78 &  35.43 &  42.30 & 0.437 & 2.098 \\ \midrule
        
        \rowcolor{gray!25}\multicolumn{6}{c}{\bf LLM-as-a-Judge: DeepSeek-V3.1} \\ \midrule

        ICE-Score       &  \underline{\textbf{65.03}} &  \underline{\textbf{55.90}} &  59.77 & 0.625 & 1.963 \\
        CodeJudge       &  61.81 &  \underline{53.26} &  44.47 & 1.025 & 1.423 \\
        CodeVisionary   &  61.45 &  51.65 &  41.89 & 1.121 & 1.327 \\
        \midrule
        
        \rowcolor{gray!25}\multicolumn{6}{c}{\bf LLM-as-a-Judge: Gemini-2.5-Flash} \\ \midrule
        
        ICE-Score       &  59.59 &  49.71 &  \underline{46.53} & 0.922 & 1.527 \\
        CodeJudge       &  60.77 &  51.95 &  34.67 & 1.293 & 1.155 \\
        CodeVisionary   &  \underline{63.22} &  \underline{53.64} &  \underline{\textbf{47.89}} & 1.002 & 1.446 \\ \midrule \midrule

        \multicolumn{6}{c}{\vspace{0.2em} \textbf{\textsc{Part II: \bench-Rectifiable}} \vspace{0.2em}} \\ \midrule
        
        \rowcolor{gray!25}\multicolumn{6}{c}{\bf Rule-Based Metrics} \\ \midrule
        
        ChrF++          &  13.25 &  10.74 &  12.54 & \underline{\textbf{0.232}} & 3.099 \\
        CodeBLEU        &  11.45 &   9.26 &  10.27 & \underline{0.238} & 2.813 \\
        Edit Similarity &  15.13 &  12.36 &  11.09 & 0.522 & 2.422 \\ \midrule
        
        \rowcolor{gray!25}\multicolumn{6}{c}{\bf LLM-as-a-Judge: DeepSeek-V3.1} \\ \midrule

        ICE-Score       &  40.06 &  33.93 &  \underline{\textbf{31.68}} & 0.752 & 2.358 \\
        CodeJudge       &  33.74 &  29.06 &   5.39 & 1.234 & 1.710 \\
        CodeVisionary   &  \underline{\textbf{49.17}} &  \underline{\textbf{40.24}} &  20.11 & 1.366 & 1.577 \\ \midrule
        
        \rowcolor{gray!25}\multicolumn{6}{c}{\bf LLM-as-a-Judge: Gemini-2.5-Flash} \\ \midrule
        
        ICE-Score       &  40.91 &  33.44 &  19.78 & 1.132 & 1.812 \\
        CodeJudge       &  33.19 &  28.08 &  -8.94 & 1.555 & 1.388 \\
        CodeVisionary   &  40.05 &  33.17 &  15.51 & 1.205 & 1.738 \\ \bottomrule
        
    \end{tabular}
    \label{tab:rq1}
    \vspace{-1.0em}
\end{table}

We measure the performance of each metric by computing the correlation between ground-truth and metric scores. The results are available in Table \ref{tab:rq1}. As we will discuss in RQ2, detecting fundamentally flawed code (0-point) is simple for modern LLMs, which inflates performance metrics. Consequently, while we report the performance on the \textbf{Full} set to represent the complete quality spectrum (establishing 0-point as the quality lower bound), we restrict our in-depth analysis to the \textbf{Rectifiable} subset (ground-truth scores of 1--5).

The results highlight that \textbf{current LLM-as-a-judge metrics are notably more effective in relative ranking than assigning absolute scores}. ICE-Score, a simple metric with one inference pass, achieves moderate rank correlations of \(\rho\approx 40\%\) for both LLMs, but its performance drops notably for absolute scoring at \(\alpha\approx 32\%\) for DeepSeek and \(\alpha\approx 20\%\) for Gemini. CodeJudge demonstrates a similar trend of near-random absolute agreement with \(|\alpha|<10\%\), while offering moderate correlations of \(\rho\approx 33\%\) for both LLMs. Likewise, while CodeVisionary can offer the highest ranking performance of \(\rho\approx 49\%\) and \(\tau\approx40\%\) for DeepSeek, this fails to translate to reliable absolute scores with \(\alpha<21\%\). This indicates that LLMs can comprehend code semantics to distinguish relative quality, but lack the internal scale calibration to match human-defined evaluation criteria, necessitating calibration before scenario-specific code evaluation.

On the more complex \textbf{Rectifiable} subset, \textbf{agentic metrics (CodeVisionary) substantially boost the ranking performance of reasoning-capable LLMs} like DeepSeek-V3.1 by 22.7\% to \(\rho\approx 49\%\), as the multi-step reasoning and tool-use procedure allows the LLM to examine the program more thoroughly and generate more accurate relative ranks. Conversely, LLMs weaker in reasoning, like Gemini-Flash, fail to benefit from the agentic workflow, maintaining a performance of \(\rho\approx 40\%\). This is likely due to their weaker internal capability, which limits their effectiveness with various tools. Nonetheless, \textbf{agentic approaches uniformly harm absolute scoring}. CodeVisionary decreases \(\alpha\) by 36.5\% for DeepSeek-V3.1 and by 21.6\% for Gemini-Flash, indicating that simple prompting is preferable when absolute scoring is prioritized. 

On the \textbf{Full} set, a converse trend emerges due to reasoning styles. DeepSeek-V3.1, being prone to lengthy reasoning chains, is encouraged by agentic approaches to deeply analyze fundamentally flawed code rather than dismissing it, leading to a 5.5\% performance degradation. Conversely, for Gemini-Flash, which prefers shorter reasoning chains, the agentic approach may act as a guardrail, forcing it to rapidly identify useless programs, resulting in a 6.1\% performance boost.

\begin{findingbox}
    \textbf{Finding 1:} Current LLM-as-a-judge metrics are capable of relatively ranking programs but lack the internal calibration to assign absolute scores. While agentic approaches substantially boost LLMs' ranking performance, they uniformly degrade absolute agreement, suggesting a fundamental trade-off between reasoning depth and absolute score calibration.
\end{findingbox}

The analysis of Earth Mover's Distance (EMD) and Mean Score exposes a systematic pessimism in existing metrics for code evaluation. The consistently lower mean scores suggest that \textbf{LLM-as-a-judge metrics exhibit profound negative biases}. The highest mean score among these metrics is observed in ICE-Score+DeepSeek-V3.1 at 2.358, which is still 19.9\% lower than the ground-truth at 2.944. Other metrics demonstrate even more severe biases at 38.5--52.9\% lower than the ground-truth. We attribute this to the objectives of RLHF training: LLMs are trained to be helpful assistants, which encourages them to actively flag defects rather than passively verify perfection. This reveals a trade-off between helpfulness (which encourages false positives to demonstrate utility) and precision (which punishes false positives).

EMDs further qualify the distribution mismatch. Surprisingly, rule-based metrics exhibit much lower EMDs at 0.232--0.522, indicating that their score distributions closely mimic the balanced distribution of ground-truth scores. However, their near-zero correlation renders this distributional fidelity useless. In contrast, LLM-as-a-judge metrics yield high EMDs at 0.752--1.555. Combined with their negative bias, we conclude that \textbf{LLM-as-a-judge metrics tend to generate biased, pessimistic scores with decent ranking correlation but poor distributional fidelity}.

\begin{findingbox}
    \textbf{Finding 2:} LLM-as-a-judge metrics exhibit systematic pessimism, frequently underestimating the quality of programs. While rule-based metrics achieve superior distributional fidelity with low correlation, LLM-as-a-judge metrics trade this fidelity for ranking capability, resulting in more accurate relative orders but biased absolute scores.
\end{findingbox}

\subsection{RQ2: Diagnostic Capability Analysis}
We further split code evaluation under our criteria into four binary classification subtasks:

\begin{itemize}
    \item \textbf{Refinability Gating} (Gating): Detects fundamentally flawed code warranting discarding. Positive: score 0; negative: scores 1--5.
    
    \item \textbf{Functionality Verification} (Func.): Detects programs that successfully implement the intended functionality. Positive: scores 3--5; negative: scores 0--2.
    
    \item \textbf{Code Quality Assessment} (Qual.): Detects high-quality code from functionally correct programs. Positive: score 5; negative: scores 3--4; excluded: scores 0--2.

    \item \textbf{Refinement Effort Estimation} (Effort): Determines whether a defective program requires structural refactoring rather than localized tweaking. Positive: scores 1 and 3; negative: scores 2 and 4; excluded: scores 0 and 5.
\end{itemize}

\begin{figure*}
    \centering
    \includegraphics[width=.8\linewidth]{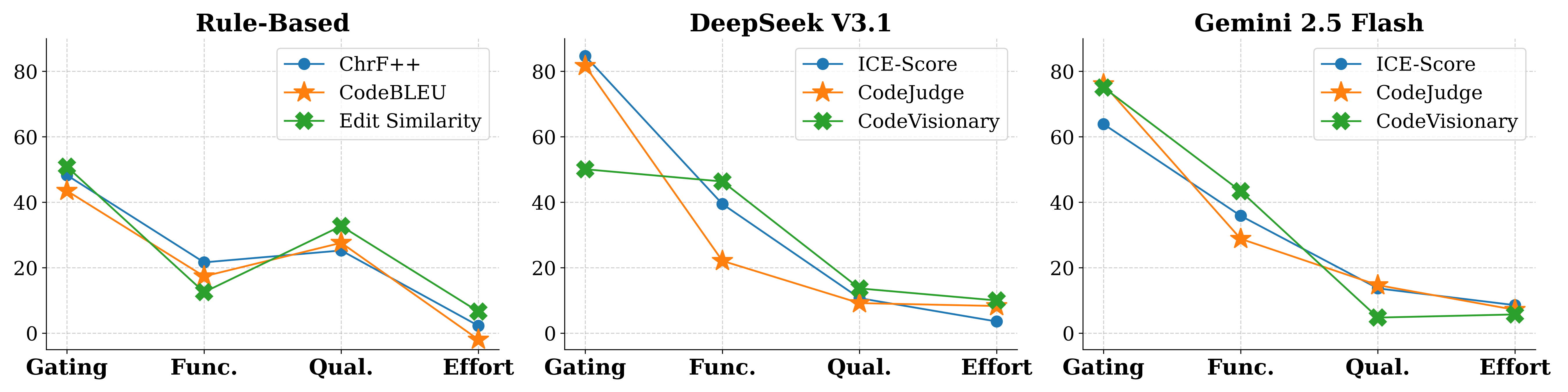}
    \caption{Plots of Matthews correlation coefficient (MCC) of each metric on different subtasks.}
    \label{fig:rq2_line}
    \vspace{-1.0em}
\end{figure*}

In Figure \ref{fig:rq2_line}, we observe that \textbf{LLM performance degrades monotonically as the task complexity increases, performing competitively on objective verification subtasks while collapsing on subjective grading ones}. Specifically, Refinability Gating is the simplest subtask, where most LLM-as-a-judge metrics achieve strong performance of MCC\(>\)75\%, as LLMs only need to recognize fundamentally flawed logic rather than identifying subtle functional defects. An exception is observed with DeepSeek-V3.1, where the inclusion of 0-point programs creates an over-thinking trap; this LLM's lengthy reasoning chains may lead to hallucinated potential in useless code, resulting in an underwhelming performance of 46\% MCC. This exception aligns with our analysis in RQ1.

Functionality Verification appears more challenging but still tractable. ICE-Score achieves a moderate 35--40\% MCC, likely because modern LLM training data includes abundant reasoning traces, enabling LLMs to internally simulate program execution. The agentic approach, CodeVisionary, provides a further 17.4\% performance boost for DeepSeek and a 21.0\% boost for Gemini-Flash. Here, the tool use steps allow them to actually execute the programs and generate reports based on concrete runtime feedback rather than probabilistic simulation.

However, current metrics are incapable of handling subjective grading subtasks, yielding disappointing performance of 5--15\% MCC on Code Quality Assessment. This requires them to examine non-functional aspects by mimicking real-world code review. Without detailed instructions or formal specifications for ``quality'', they may be tempted to report superficial characteristics like lack of comments, rather than distinguishing nuanced structural flaws. On Refinement Effort Estimation, no metrics exceed 10\% MCC. This fundamentally different subtask requires LLMs to not only detect defects but also simulate repair trajectories to estimate the cost. Our results indicate that while LLMs can detect code defects, they struggle to estimate the magnitude of repair effort.

\begin{findingbox}
    \textbf{Finding 3:} LLM performance correlates negatively with task complexity. Objectively verifiable evaluation tasks like functionality verification prove tractable. Subjective grading tasks like code quality assessment remain unmanageable.
\end{findingbox}

Among verifiable subtasks, LLM-as-a-judge (\(\text{MCC}>75\)\% in most cases) remarkably outperforms rule-based metrics (MCC\(<\)51\%) on Refinability Gating. While the agentic approach improves Gemini-Flash's performance, it paradoxically degrades DeepSeek-V3.1, reinforcing the ``over-thinking'' hypothesis where unnecessarily lengthy reasoning introduces noise into simple decisions. On the more challenging Functionality Verification subtask, the performance gap persists, as rule-based metrics struggle to capture runtime logic, and LLM-as-a-judge metrics maintain moderate competence. Notably, the agentic CodeVisionary consistently boosts performance by 17--21\%, indicating that at this moderate difficulty, the overhead of agentic usage pays off by runtime context and deeper reasoning. Conversely, rule-based metrics achieve superior performance on Code Quality Assessment at 25--33\% MCC compared to LLMs, which struggle to offer reliable non-functionality assessment. This demonstrates that such conventional, rule-based metrics remain the most robust method in code quality assessment. Finally, no metrics are capable of correctly estimating refinement effort at near-zero MCC, highlighting a critical gap in automated maintenance estimation.

\begin{findingbox}
    \textbf{Finding 4:} LLM-as-a-judge dominates objectively verifiable subtasks, while agentic approaches further boost performance. However, rule-based metrics remain superior for non-behavioral evaluation. Both paradigms cannot estimate refinement effort reliably.
\end{findingbox}

\subsection{RQ3: Scoring Consistency}
\begin{table}[]
    \centering
    \caption{Metric consistency. \textbf{Run} metrics are computed across five runs to measure sample-level consistency. \textbf{GT} metrics are computed for each run against the ground-truth scores to measure variance in dataset-level performance.}
    \begin{tabular}{c|cc|cc}
    \toprule
    
    \multirow{2}{*}{\bf Metric} & \multicolumn{2}{c|}{\bf Consistency} & \multicolumn{2}{c}{\bf Performance} (Mean ± Std.) \\
    & \(\alpha_\text{Run}\) & ICC\(_\text{Run}\) & \(\rho_\text{GT}\) & \(\alpha_\text{GT}\) \\ \midrule

    \rowcolor{gray!40}\multicolumn{5}{c}{\bf LLM-as-a-Judge: Gemini-2.5-Flash} \\ \midrule

    ICE-Score      &  \underline{\bf 86.61} &  \underline{\bf 85.23} &  \underline{\bf 39.92} ±   2.40 &  \underline{\bf 16.84} ±   2.57 \\
    CodeJudge      &  73.89 &  77.89 &  29.82 ±   \underline{\bf 1.42} & -17.71 ±   \underline{\bf 2.17} \\
    CodeVisionary  &  51.66 &  52.09 &  34.36 ±   1.66 &   3.25 ±   3.66 \\ \bottomrule

    \end{tabular}
    \label{tab:rq3}
    \vspace{-1.0em}
\end{table}

We study metric consistencies with Gemini-2.5-Flash in Table \ref{tab:rq3}. Each metric is executed five times on each program.

Both simpler metrics, ICE-Score and CodeJudge, show high consistency for sample-level scores with \(\alpha>73\%\) and \(\text{ICC}>77\%\), indicating that prompting the LLM once provides surprisingly consistent results on a single sample. In contrast, the complex agentic metric, CodeVisionary, behaves notably more inconsistently with 40.4\% lower \(\alpha\) and 38.9\% lower ICC. This can be attributed to accumulated uncertainty: Each step during CodeVisionary's pipeline of action planning and tool calling introduces stochasticity, leading to higher variance in the final scores. The other metrics only include 1--2 LLM invocations, minimizing this uncertainty accumulation. For the dataset-level performance, the standard deviations of \(\rho\) and \(\alpha\) stay consistently below \(4\%\), indicating that despite stochasticity during LLM invocation, the aggregate dataset-level performance impact is generally marginal.

\begin{findingbox}
    {\bf Finding 5:} Procedural complexity mainly affects sample-level scoring consistency. Simpler LLM-as-a-judge metrics with few LLM invocations assign highly consistent scores to the same program across runs. In contrast, complex agentic workflows accumulate stochasticity across intermediate states, leading to larger variance in sample-level scores. However, when aggregated over the benchmark, each run achieves similar overall performance, suggesting that such instability has limited impact on dataset-level conclusions.
\end{findingbox}

\section{Discussion\label{sec:dis}}

\subsection{Illustrative Example}
\begin{figure}
    \centering
    \includegraphics[width=0.45\textwidth]{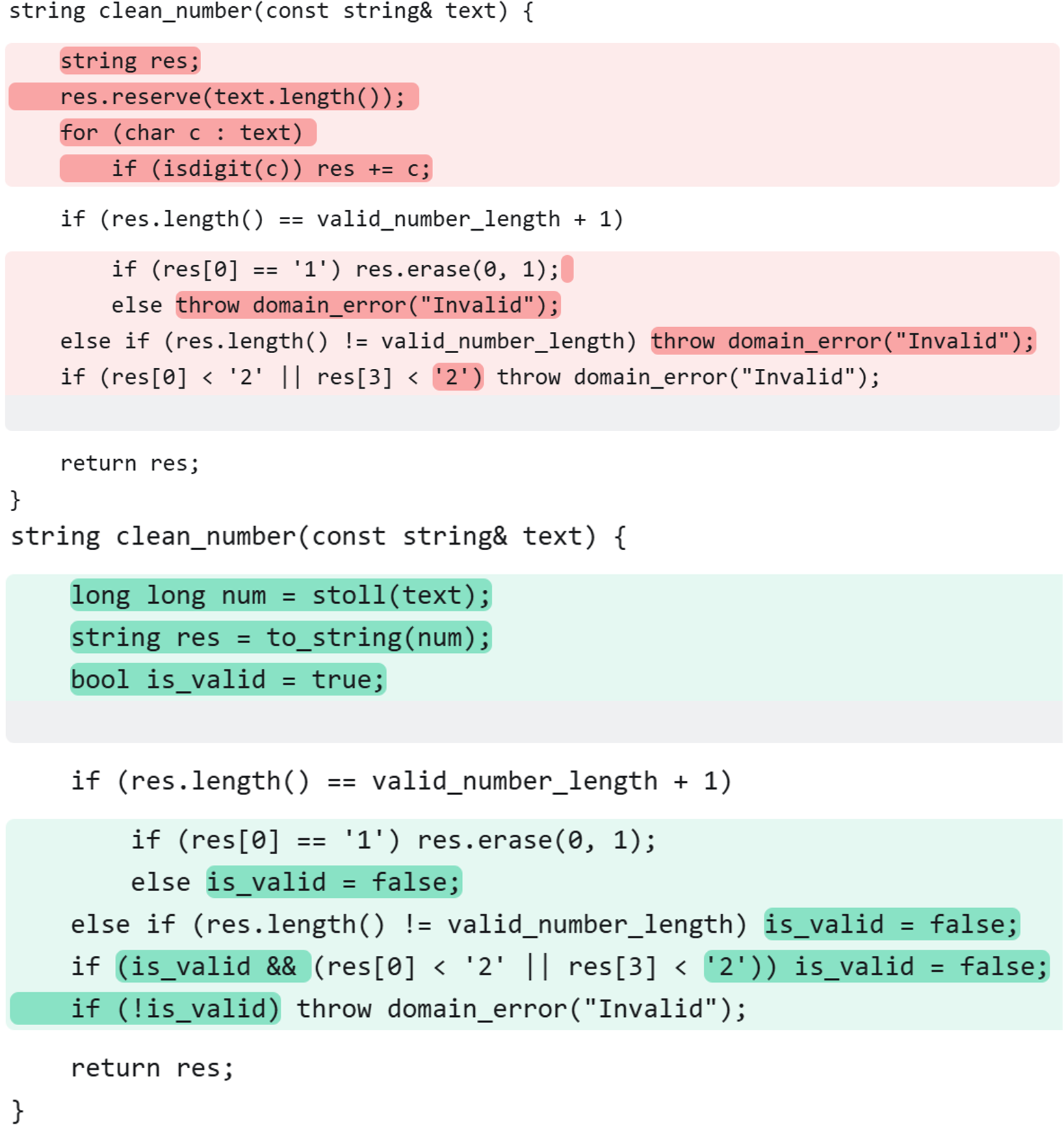}
    \caption{Illustrative example in \bench.}
    \label{fig:case}
    \vspace{-1.0em}
\end{figure}

Figure \ref{fig:case} illustrates one \bench example. The program above is 5-point and production-ready. \tool first introduces a semantics-preserving intermediate variable \(is\_valid\), and then adds a localized functional defect by converting \(text\) to \(num\) before filtering non-digit characters. The perturbed program below preserves the original structure but fails relevant tests. Since the functional defect can be repaired by a localized edit, it is considered 2 points.

\subsection{Implications of Study}

\subsubsection{For Developers}

\begin{enumerate}
    \item \textbf{Prefer ranking over absolute gating.} Current LLM-as-a-judge metrics are competitive at ranking candidates but suffer from severe systematic bias in absolute scoring. Developers should avoid using raw LLM scores for quality gating, for instance, accepting the code if score\(>\)3, unless the metric has been strictly calibrated.

    \item \textbf{Delineate verifiable and subjective evaluation tasks.} LLMs are more reliable in verifiable evaluation tasks like functionality assessment. However, they underperform in subjective grading tasks, like maintainability review, in which developers should treat LLM feedback as diagnostic rather than definitive.

    \item \textbf{Default to simple methods for stability.} Contrary to the assumption that agents consistently improve performance, simple prompting often yields higher scoring stability and efficiency for standard evaluation. Agentic approaches should be reserved for complex evaluation tasks with proper variance reduction methods.
\end{enumerate}

\subsubsection{For Researchers}

\begin{enumerate}
    % \item \textbf{Absolute agreement is essential.} Rank correlation coefficients like \(\rho\) is insufficient to meta-evaluate the reliability of metrics. Researchers should report absolute agreement meta-metrics like \(\alpha\) to expose systematic biases, and design measures to combat such biases.

    \item \textbf{Adopt quality-constrained benchmark construction.} Previous code evaluation benchmarks often rely on unconstrained LLM generation, inevitably skewing toward high-quality code, underrepresenting the lower-quality spectrum. To reliably test the metrics' discriminative capability across the entire quality spectrum, future work should actively constrain the quality of generated code to guarantee a balanced distribution.

    \item \textbf{Address the subjectivity alignment gap.} While LLMs are already competitive at detecting verifiable correctness, future work should pivot towards the harder evaluation task of subjective quality grading and target the challenging problem of non-behavioral code evaluation.

    \item \textbf{Prioritize absolute agreement and metric stability.} Evaluating metrics solely via rank correlation obscures systematic biases. Researchers should also report absolute agreement meta-metrics like \(\alpha\) to expose the biases. Furthermore, certain metrics like agentic methods amplify stochasticity and degrade scoring consistency. Therefore, future work should focus on variance reduction to stabilize agentic judges while preserving their reasoning depth.
    % \item \textbf{Stabilize agentic approaches.} Our findings reveal that deep tool use and reasoning chains introduce substantial stochasticity, degrading metric consistency. Future work should stabilize agentic judges to reconcile the trade-off between action depth and evaluation reliability.
\end{enumerate}

\subsection{Threats to Validity}
\begin{itemize}
    % \item {\bf Programming Languages.} Our benchmark covers three prevalent, multi-paradigm languages, namely C++, Python, and Java. Consequently, our findings may not directly generalize to languages with distinct paradigms or domains, such as Rust for system programming and JavaScript for web development. We mitigate this by selecting languages representing a significant portion of modern software development, and leave extensions to other languages as future work.

    % \item {\bf Usage of Synthetic Data.} \bench relies on synthetically perturbed programs, which may not perfectly reflect the distribution of real-world programming errors. We adopt this approach because open-source repositories are overwhelmingly skewed towards high-quality code, hindering the construction of a balanced evaluation benchmark. We mitigate this potential artificiality through realistic perturbation rules designed to mimic real-world errors; and the \stageb stage, where all perturbations and scores are manually verified to represent meaningful, realistic defects with accurate scores.

    \item {\bf Domain Generalizability.} \bench mainly features algorithmic challenges and library usage tasks across three prevalent languages (C++, Java, Python). The findings may thus not fully generalize to languages with distinct paradigms like Rust or projects with repository-level dependencies. We mitigate this by including data sources with cross-file dependencies like Aider-Polyglot, and making perturbation rules inherently language- and dependency-agnostic. This allows \tool to readily extend to more complex, domain-specific codebases.

    \item {\bf Data Contamination.} LLMs meta-evaluated as judges may have memorized parts of the reference programs during training, potentially inflating meta-evaluation results. We mitigate this by including data sources with continuous updates like LiveCodeBench. Furthermore, our reliance on perturbed programs naturally mitigates this risk, because they are uniquely generated with novel flaws absent from public sources like GitHub.

    \item {\bf Subjectivity in Criteria Interpretation.} Our evaluation criteria rely on human distinction between minor tweaks and major refactors, potentially introducing subjectivity in our ground-truth labels. We mitigate this with \stagebb, providing the annotator with a wide range of objective evidence, like static analysis results and \texttt{diff}. This reframes the annotating task from open-ended scoring to a more objective verification. We also conduct a preliminary study as specified in Section \ref{sec:prelim}, and the high inter-rater agreement indicates high objectivity during manual annotation.
\end{itemize}
\section{Conclusion\label{sec:con}}
In this paper, we propose \tool, a framework to synthesize code evaluation data to meta-evaluate existing LLM-as-a-judge metrics for code. \tool involves a \stagea stage to synthesize programs and a \stageb stage to calibrate scores, addressing issues of previous benchmarks. We use this framework to craft a balanced, verified benchmark, \bench, conduct experiments with multiple recent LLM-as-a-judge metrics, and release our data\footnote{\url{https://github.com/BackOnTruck/axiom-llm-judge}.} and findings.

% \begin{acks}
% WIP: acknowledgments
% \end{acks}

\bibliographystyle{IEEEtran}
\bibliography{references}

% \appendix

\end{document}